\documentclass[a4paper,twocolumn,10pt]{article}

\usepackage[margin=0.75in]{geometry}

\pdfoutput=1

\usepackage{graphicx}
\usepackage{bm}
\usepackage{dcolumn}
\usepackage{setspace}
\usepackage{abstract}
\usepackage{multirow}
\usepackage{multicol}
\usepackage{sidecap}
\usepackage{caption}
\usepackage{subcaption}

\usepackage[affil-it]{authblk}

\usepackage[utf8]{inputenc}
\usepackage{lineno,hyperref}
\usepackage{cite}
\usepackage{color,soul}

\usepackage{balance}

\hypersetup{colorlinks=true, linkcolor=blue, citecolor=red}

\begin{document}
\title{\bf \Large Twinning to slip transition in ultrathin BCC Fe nanowires}

\date{}

\author{G. Sainath\footnote{email : sg@igcar.gov.in}, B.K. Choudhary}

\affil { Deformation and Damage Modeling Section, Materials Development and Technology Division\\ 
Indira Gandhi Centre for Atomic Research, HBNI, Kalpakkam 603102, Tamil Nadu, India}
\twocolumn[

\maketitle

\begin{onecolabstract}


\baselineskip 13pt

We report twinning to slip transition with decreasing size and increasing temperature in ultrathin
$<$100$>$ BCC Fe nanowires. Molecular dynamics simulations have been performed on different nanowire
size in the range 0.404-3.634 nm at temperatures ranging from 10 to 900 K. The results indicate that 
slip mode dominates at low sizes and high temperatures, while deformation twinning is promoted at high
sizes and low temperatures. The temperature, at which the nanowires show twinning to slip transition,
increases with increasing size. The different modes of deformation are also reflected appropriately 
in the respective stress-strain behaviour of the nanowires. \\ 



\end{onecolabstract}

]
\renewcommand{\thefootnote}{\fnsymbol{footnote}} \footnotetext{* email : sg@igcar.gov.in}

{\small

\baselineskip 13pt

\section{\large Introduction}

Dislocation slip and deformation twinning are two important deformation mechanisms in crystalline materials. 
In general, plastic deformation occurs by the slip of dislocations at room temperature and conventional strain 
rates, while twinning is observed in conditions that lead to high stresses such as high strain rates
and/or low temperatures. However, it is now well established that the deformation behaviour of single crystals 
with dimensions reduced to nanoscale is qualitatively different from their bulk counterparts \cite{Greer-PMS}. 
At nanoscale, the size in addition to temperature and strain rate also plays an important role on the deformation 
behaviour \cite{Wei-Cai-JMC}. In this regard, understanding the size dependence of deformation behaviour in 
metallic nanowires becomes important and hence attracted huge interest among the materials community.

\vspace{0.2cm}

Experimental and molecular dynamics (MD) simulation studies have shown that the size influences the elastic 
modulus \cite{Liang}, yield and flow stresses \cite{Uchic}, dislocation nucleation and character 
\cite{cross-over-NL,Ag-transition}, deformation mechanisms \cite{cross-over-NL,Ag-transition,Oh,Sedlmayr,Roos} 
and failure behaviour \cite{Small}. With decreasing size, a transition in deformation mechanisms from dislocation 
slip to twinning has been reported in many FCC nanowires \cite{cross-over-NL,Ag-transition,Oh,Sedlmayr,Roos}. 
For example, a transition from full dislocation slip to twinning/partial dislocations has been demonstrated in 
Cu nanowires for sizes ranging from 70 to 1000 nm \cite{cross-over-NL}. In Cu nanowires of size higher than 
150 nm, deformation remains dominated by the slip of full dislocations, while partial dislocations/twinning has 
been observed below 150 nm \cite{cross-over-NL}. In Ag nanowires of size 11 nm and above, dominance of slip 
of full dislocations has been observed, while stacking faults and twins governs plastic deformation in 5-8 nm 
nanowires \cite{Ag-transition}. Interestingly, with further decrease in size to less than 3 nm, plastic 
deformation accommodated by relative slip between two adjacent \{111\} planes without any dislocations has been 
reported \cite{Ag-transition}. A similar transition has also been observed in Au thin films, nanowires and 
nanopillers \cite{Oh,Sedlmayr,Roos}. These results suggest that there exist a clear transition in deformation
mechanisms in FCC nanowires. Contrary to this, Yu et al. \cite{HCP-Size} demonstrated a reverse transition 
from twinning at higher sizes to dislocation slip at smaller sizes in HCP Ti. Using in-situ experiments, deformation 
twinning has been observed in HCP Ti single crystal of size 1 $\mu m$ and above \cite{HCP-Size}. Below this 
size, twinning is entirely replaced by the slip of dislocations \cite{HCP-Size}. These observations strongly 
suggest that there is a size limit below which twinning does not occur in HCP Ti. In this context, it is important 
to understand how the size influences the deformation behaviour in BCC nanowires. 

\vspace{0.2cm}

In the past, many MD simulations 
have been carried out to understand the deformation mechanisms in BCC Fe and Mo nanowires 
\cite{Mo,Healy,Sainath-Orientation,Adutta,Hagen,Sainath-100-size}. All these studies have concluded that the 
deformation in $<$100$>$ BCC Fe and Mo nanowires occurs by deformation twinning. The twinning in BCC nanowires 
is confirmed by the recent experimental observation in $<$100$>$ W nanowires of size 15 nm \cite{Nature-exp}. 
The Schmid factor calculations also predict twinning in $<$100$>$ BCC crystals \cite{Nature-exp}. 
However, here we show that, when the nanowire size is reduced below 3.23 nm, the twinning in $<$100$>$/\{110\} 
BCC Fe nanowires is completely replaced by dislocation slip at high temperatures. MD simulations have been carried 
out on nanowires with cross section width (d) ranging from 0.404 to 3.634 nm for temperatures in the range 
10-900 K. The combined influence of size and temperature twinning to slip transition for ultra-thin size along 
with stress-strain behaviour of $<$100$>$ BCC Fe nanowires has been presented.

\section{\large Simulation Details}

Molecular dynamics (MD) simulations have been performed using LAMMPS package \cite{LAMMPS} employing an embedded 
atom method (EAM) potential for BCC Fe given by Mendelev and co-workers \cite{Mendelev}. Earlier several studies 
have shown that this potential appropriately describes the deformation behaviour of BCC Fe nanowires 
\cite{Healy,Sainath-Orientation,Adutta,Sainath-100-size}. 

\vspace{0.2cm}

BCC Fe nanowires oriented in $<$100$>$ axial direction
with \{110\} as side surfaces have been considered in this study. MD Simulations have been performed on nanowires 
with cross section width (d) ranging from 0.404 to 3.634 nm. In all the nanowires, the length (l) was twice the 
cross section width (d). Periodic boundary conditions have been chosen along the nanowire length direction, while 
the other directions were kept free in order to mimic an infinitely long nanowire. After the initial construction 
of nanowire, energy minimization was performed by conjugate gradient method to obtain a stable structure. To put 
the sample at the required temperature, all the atoms have been assigned initial velocities according to the 
Gaussian distribution. Following this, the nanowire system was thermally equilibrated to a required temperature 
for 125 ps in canonical ensemble using Nose-Hoover thermostat. The Velocity Verlet algorithm was used to integrate 
the equations of motion with a time step of 5 fs. 

\vspace{0.2cm}

Following thermal equilibration, the tensile deformation was 
performed at a constant strain rate of $1\times$ $10^8$ s$^{-1}$ along the axis of the nanowire. For each cross 
section width (d) of nanowires, MD simulations have been performed at different temperatures ranging from 10 K 
to 900 K. Further, at each size and temperature conditions, five independent MD simulations with different 
random number seeds have been performed to make statistically meaningful conclusions. The stress was calculated 
from the Virial definition of stress \cite{Virial-Yip,Virial-Zhou}. The visualization of atomic configurations 
was performed using AtomEye\cite{AtomEye} and OVITO \cite{OVITO}.

\section{\large Results and Discussion}

Figure \ref{stress-strain} shows stress-strain behaviour of $<$100$>$/\{110\} BCC Fe nanowires with d = 1.615 nm at 
different temperatures ranging from 10 to 500 K. It can be seen that at all temperatures, the nanowires undergo an 
initial elastic deformation up to a peak stress followed by an abrupt drop in flow stress due to yielding. Following 
yielding, the stress-strain behaviour of the nanowires depends strongly on the temperature. At low temperatures of 
10 and 100 K, the stress-strain behaviour during plastic deformation exhibits uniform flow stress oscillations about 
a constant mean value up to a strain of 0.4 (Figure \ref{stress-strain}a). Following this, the nanowires display second 
elastic peak followed by stress drop and continuous decrease in stress till failure. On the other hand, at high 
temperatures i.e. at 200 K and above, the nanowires don’t show any second elastic peak (Figure \ref{stress-strain}b). 
During plastic deformation, a continuous decrease in stress with large fluctuations (jerky flow) till failure can be 
seen (Figure \ref{stress-strain}b). The observed significant difference in the stress-strain behaviour clearly 
suggests that different deformation mechanisms are operative at low (i.e., 10 and 100 K) and high (i.e., at 200 K and 
above) temperatures during tensile deformation in $<$100$>$ BCC Fe nanowires.

\begin{figure}
\centering

\begin{subfigure}[b]{0.47\textwidth}
\includegraphics[width=\textwidth]{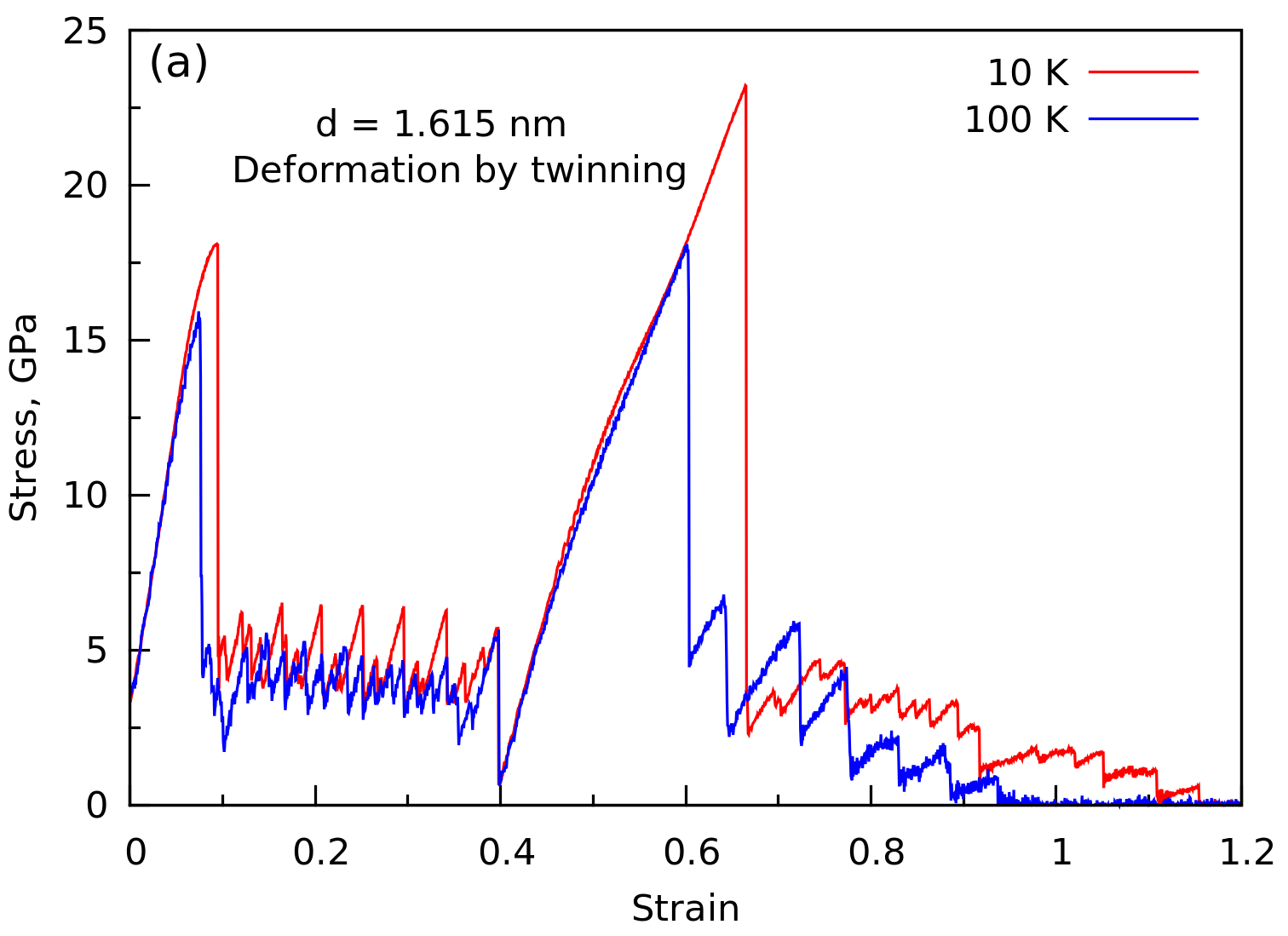}
\end{subfigure}
\qquad
\begin{subfigure}[b]{0.47\textwidth}
\includegraphics[width=\textwidth]{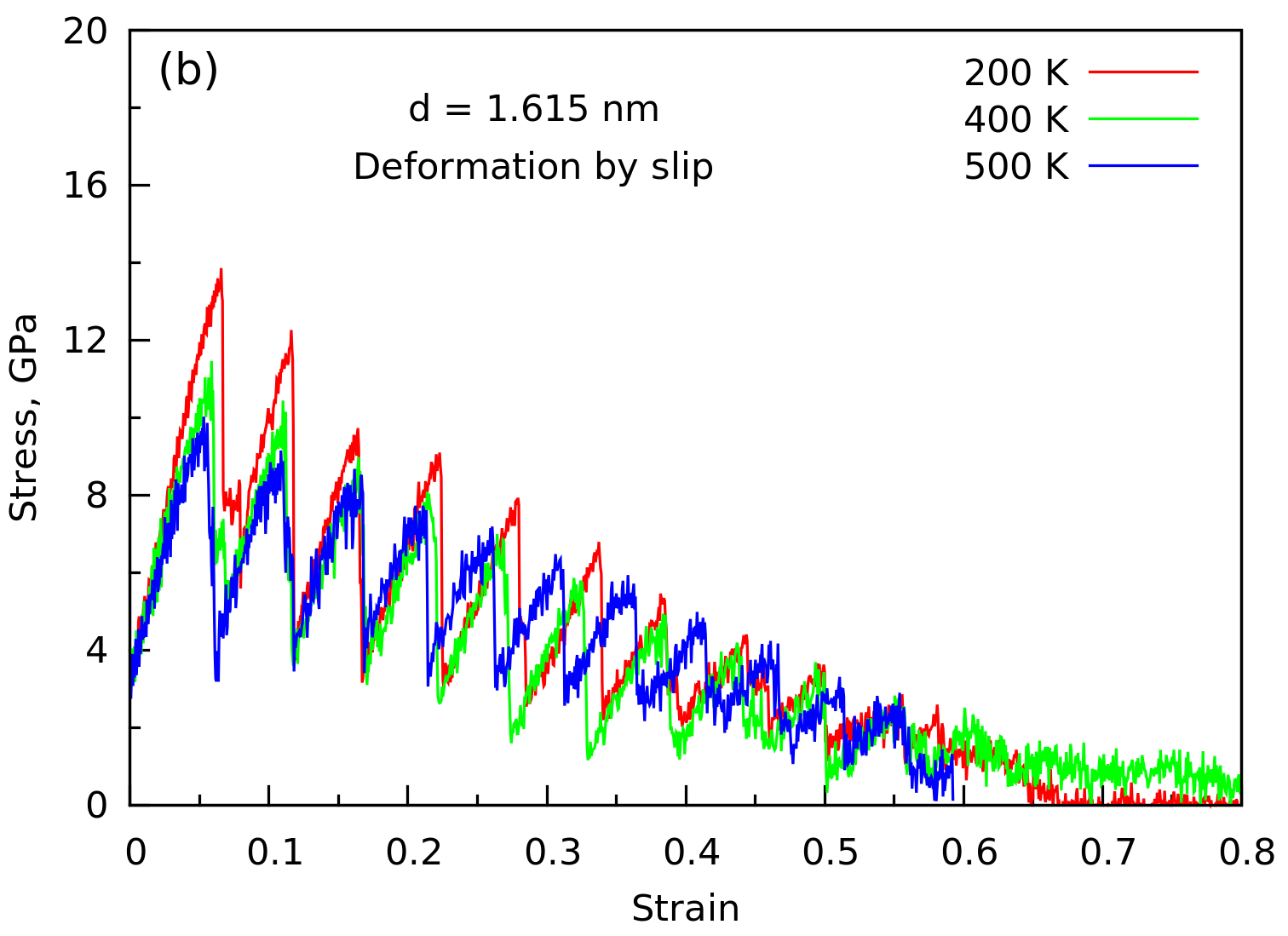}
\end{subfigure}
 \caption {\small The stress-strain behaviour of $<$100$>$/\{110\} BCC Fe nanowires at (a) low (10-100 K) and 
 (b) high (200-500 K) temperatures.} 
 \label{stress-strain}
 \end{figure}
 
 \vspace{0.2cm}
 
In order to understand the difference in stress-strain behaviour, the atomic configurations have been analysed as a 
function of strain at various temperatures. Figure \ref{Twinning} shows the deformation behaviour in nanowire with 
d = 1.615 nm at the lowest temperature of 10 K. It can be seen that initially perfect nanowire (Figure \ref{Twinning}a) 
yields by the nucleation of a twin embryo, associated with an abrupt drop in flow stress. Following this, the twin 
embryo becomes a full twin enclosed by two twin boundaries (Figure \ref{Twinning}b). With increasing strain, the twin 
grows along the axis of the nanowire (Figure \ref{Twinning}c) and progressively reorients the nanowire within the 
twinned region. The growth of twin by repeated nucleation and glide of 1/6$<$111$>$ twinning partials on the twin 
boundaries is responsible for the constant but fluctuating flow stress about a mean value in the strain range 0.1-0.4 
\cite{Sainath-100-size}. Once the twin completely sweeps the $<$100$>$/\{110\} nanowires, orientation of the nanowire 
changes to $<$110$>$ tensile axis having \{100\} and \{110\} as side surfaces (Figure \ref{Twinning}d). It can be clearly 
seen that the new $<$110$>$ reoriented nanowire is completely defect free (Figure \ref{Twinning}d) and with increasing 
deformation, it undergoes an elastic deformation once again as reflected in the occurrence of second elastic peak in 
the stress-strain curve (Figure \ref{stress-strain}a). Following the second elastic deformation, the reoriented nanowire
deforms by full dislocation slip leading to neck formation and final failure (Figure \ref{Twinning}e) at relatively higher 
values of strain. Similar stress-strain behaviour and deformation dominated by twinning mechanism have been observed 
up to 150 K.

\begin{figure}
\centering
\includegraphics[width=8cm]{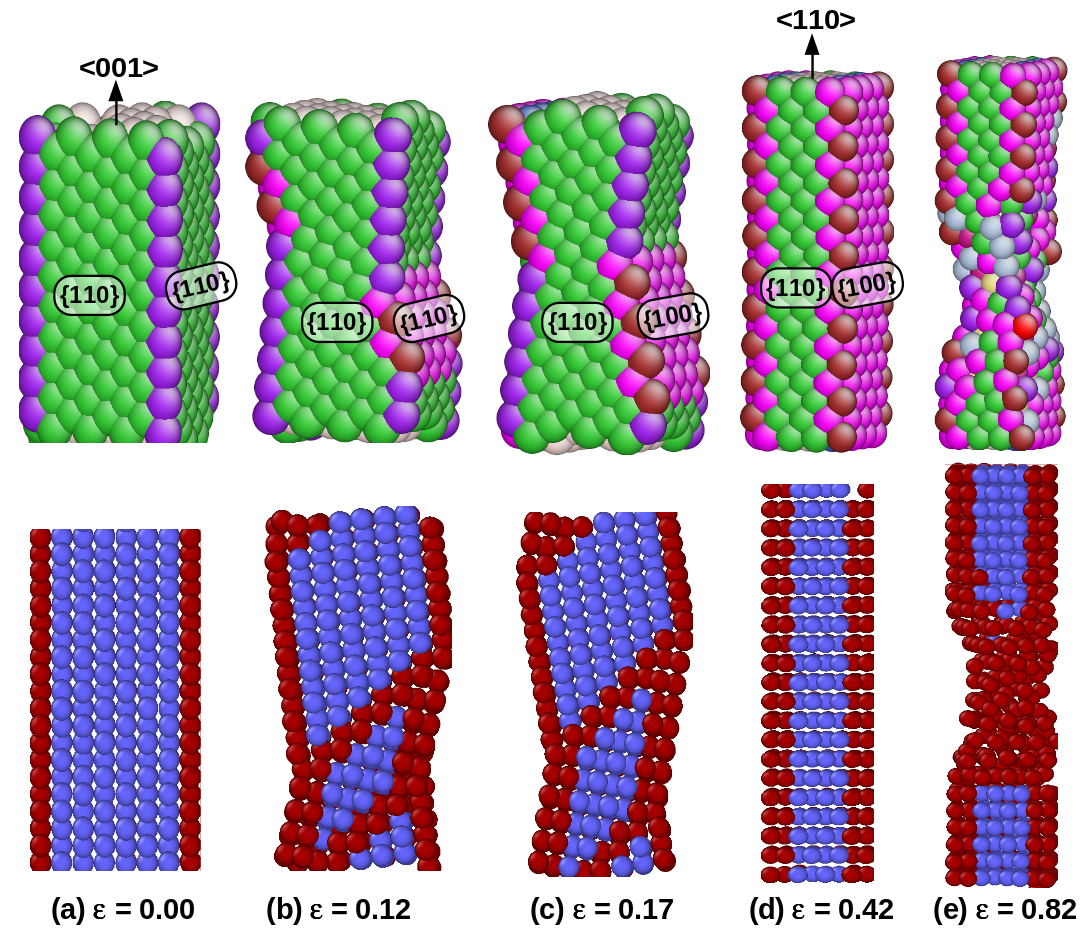}
\caption {\small The typical deformation behaviour by twinning mechanism under the tensile loading of $<$100$>$/\{110\} 
BCC Fe nanowire with d = 1.615 nm at the lowest temperature of 10 K. The snapshots in top row are coloured according 
to the atom’s coordination number and indicates the changes in surface orientation due to twinning. In bottom row, the 
colour is according to the common neighbour analysis and it clearly shows the presence of twin boundaries.}
\label{Twinning}
\end{figure}

\vspace{0.2cm}

Contrary to deformation by twinning at low temperatures, deformation dominated by dislocation slip at relatively higher 
temperatures (200 K and above) is shown for 300 K in Figure \ref{Slip} as an example. The initially perfect nanowire 
(Figure \ref{Slip}a) yields by the nucleation of 1/2$<$111$>$ full dislocations and with increasing strain, dislocations 
glide on their respective slip planes and finally annihilate at the opposite surface. As a result of deformation by 
full dislocation slip, the slip steps have been observed on the surface of the nanowire (Figure \ref{Slip}b-c). The 
absence of surface reorientation clearly indicates that no twinning occurred at 200 K and above. The discrete events 
of nucleation, glide and annihilation of dislocations is responsible for the observed jerky flow in the stress-strain 
curve (Figure \ref{stress-strain}b). Following plastic deformation, the nanowire fails by shear along the dominant 
slip plane at relatively lower values of strain to failure (Figure \ref{Slip}d). These results clearly suggest that the 
BCC Fe nanowire undergoes twinning to slip transition with increasing temperature. The different modes of deformation 
are also reflected in the respective stress-strain behaviour of the nanowires (Figure \ref{stress-strain}). In case of 
twinning mode of deformation, a plateau in the flow stress is generally observed, and the occurrence of second elastic 
peak necessarily suggests twinning followed by reorientation \cite{Mo,Sainath-100-size}. On the other hand, 
deformation dominated by full dislocation slip results in continuous decrease in flow stress along with jerky flow 
\cite{Mo}.

\begin{figure}[h]
\centering
\includegraphics[width=8cm]{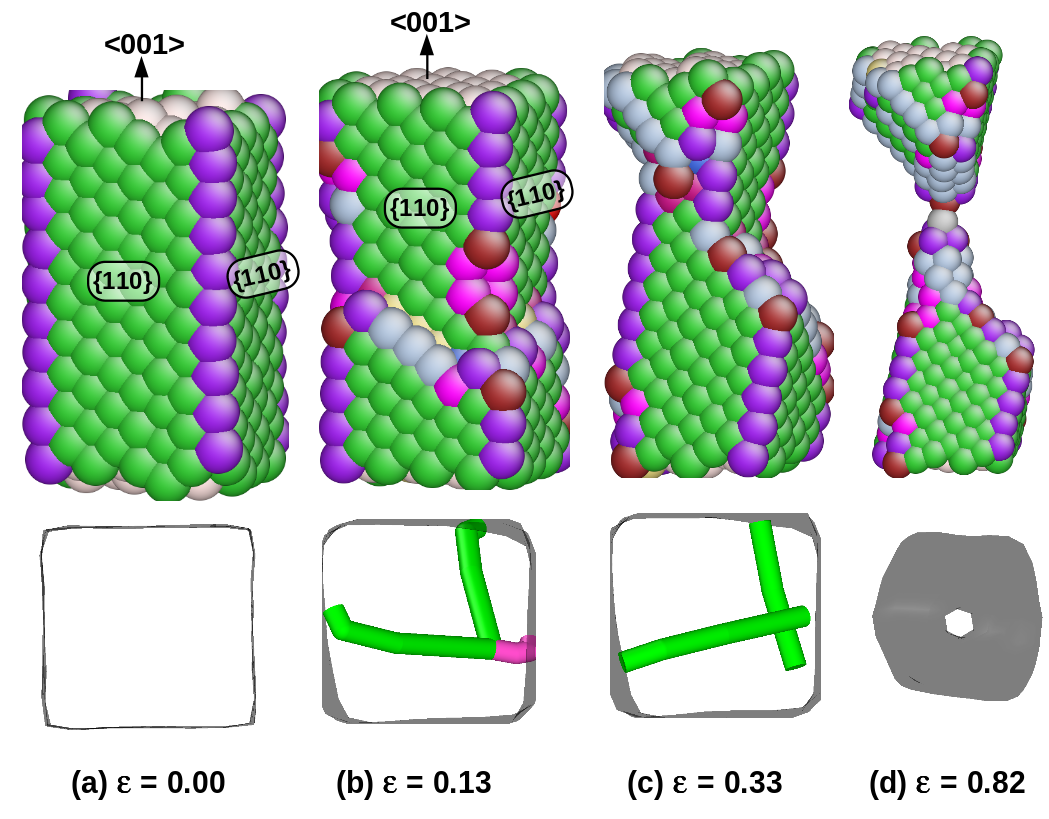}
\caption {\small The deformation behaviour by dislocation slip under tensile loading of $<$100$>$/\{110\} BCC Fe 
nanowire with d = 1.615 nm at 300 K. The atoms are coloured according to their coordination number. It can be seen 
that the surface orientation remains same till failure. The bottom row is the view along the nanowire axis showing 
the 1/2$<$111$>$ (green lines) and $<$100$>$ dislocations (magenta lines).}
\label{Slip}
\end{figure}


In order to demonstrate the combined influence of size and temperature on twinning to slip transition in $<$100$>$ BCC 
Fe nanowires, the results obtained for different sizes in the range 0.404 to 3.634 nm and for temperatures ranging 
from 10 to 900 K are shown in Figure \ref{Figure4}a. The deformation mechanisms map separating the two different regions 
of twinning and slip modes of deformation with respect to size and temperature are marked in Figure \ref{Figure4}a. It 
can be clearly seen that the temperature at which the nanowires show twinning to slip transition increases with increase
in nanowire size. In other words, at each temperature, there is a critical size below which twinning cannot occur. At 
low sizes and high temperatures, the slip mode dominates, while at high sizes and low temperatures, deformation twinning 
is promoted. In the lowest nanowire size of d = 0.404 nm, even though the deformation by slip is observed, but dislocations 
have not been noticed. Instead of dislocations, the relative slip between the two adjacent slip planes has been seen. In 
all other sizes, where slip dominates, 1/2$<$111$>$ full dislocations have been observed as shown in the bottom row in 
Figure \ref{Slip}. For nanowire size of 3.23 nm and above, twinning to slip transition has not been observed, and deformation
by twinning has been noticed irrespective of temperature. This is also reflected in the flow stress plateaus and occurrence 
of second elastic peak following reorientation in stress-strain behaviour typically shown for d = 3.634 nm for the 
temperature range 10-900 K in Figure \ref{Figure4}b. The analysis of atomic configurations for nanowire size with d = 
3.634 nm also indicated the dominance of twinning at all temperatures examined. Healy and Ackland \cite{Healy} have reported
dominance of twinning in $<$100$>$ BCC Fe nanopillar with relatively higher size of 5.8 nm at 300 K. It is important to 
mention that twinning to slip or slip to twinning transition has not been observed under compressive loading in $<$100$>$ 
BCC Fe nanowires, and the deformation remains dominated by dislocation slip for the size and temperature conditions shown 
in Figure \ref{Figure4}a.

\begin{figure}
\centering

\begin{subfigure}[b]{0.45\textwidth}
\includegraphics[width=\textwidth]{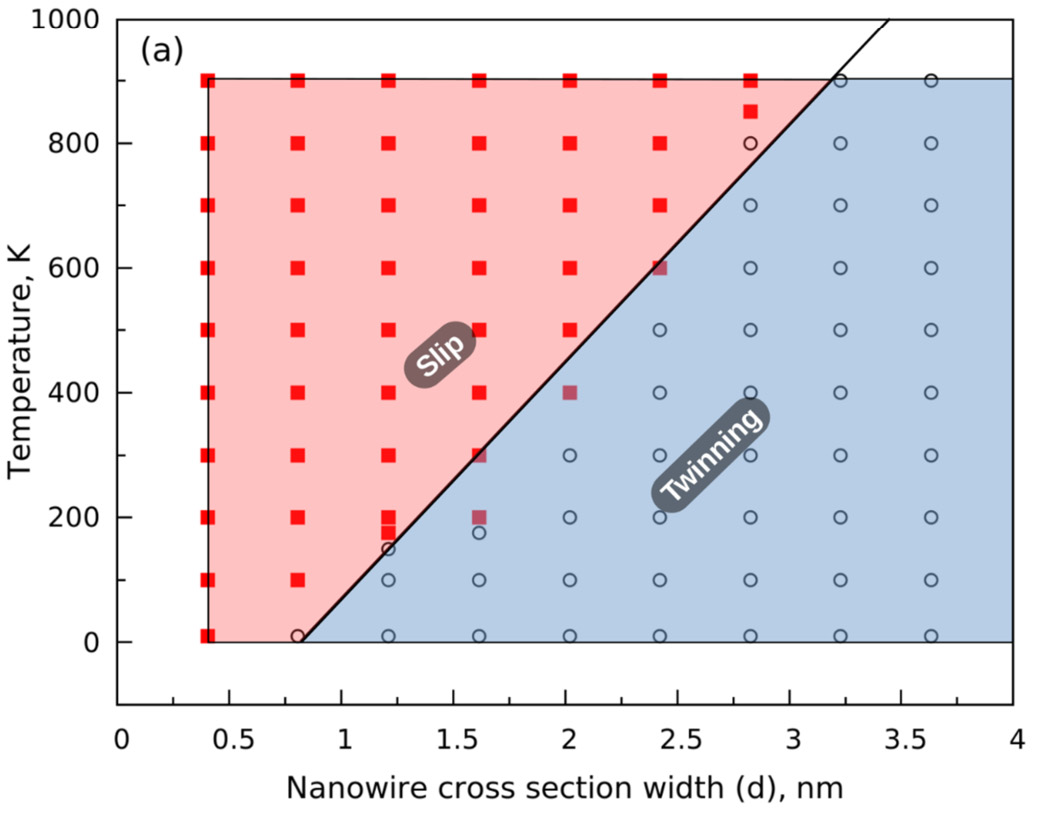}
\end{subfigure}
\qquad
\begin{subfigure}[b]{0.45\textwidth}
\includegraphics[width=\textwidth]{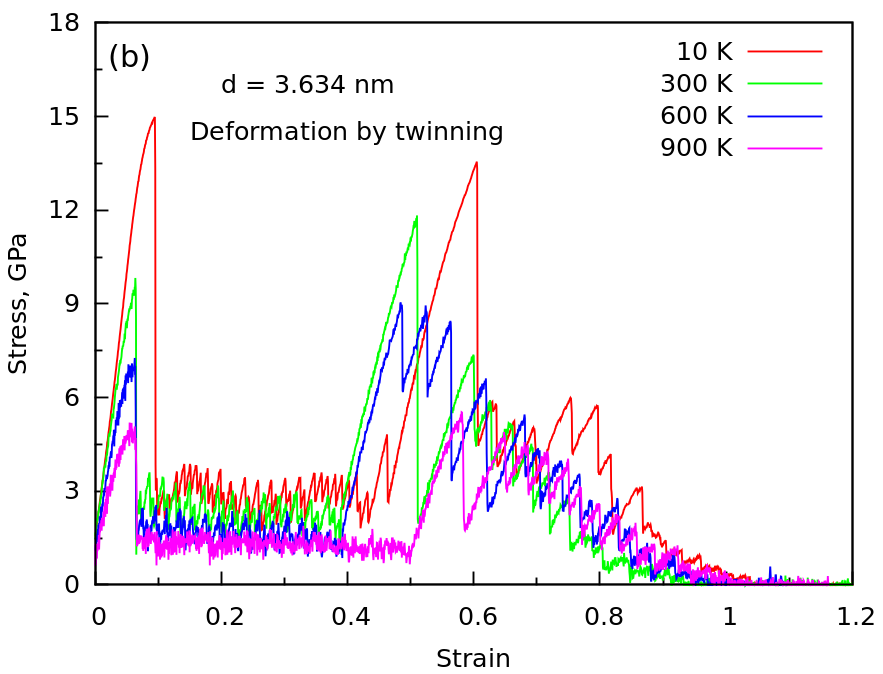}
\end{subfigure}
 \caption {\small (a) The deformation mechanisms map showing the regions dominated by twinning and slip with respect to 
nanowires size and temperature, and (b) stress-strain behaviour of the nanowire with d = 3.634 nm at different temperatures 
in the range 10-900 K.} 
\label{Figure4}
\end{figure}
 
\vspace{0.2cm}

Recent large scale MD simulation studies have shown that the deformation by twinning or dislocation slip in BCC crystals 
depends on many factors such as strain rate, temperature and initial microstructure \cite{Bulatov}. It has been shown 
that in samples free of initial defects, the twinning is inherently favoured at low temperatures and high strain rates 
\cite{Bulatov}. However, the presence of initial dislocations may change the deformation mode from twinning to 
dislocation slip \cite{Bulatov,Sai-PML}. The occurrence of twinning under tensile loading in $<$100$>$ BCC Fe nanowires 
is in agreement with earlier MD simulation studies \cite{Healy,Sainath-Orientation,Sainath-100-size} and experimental 
observations on BCC W nanowires of size 15 nm \cite{Nature-exp}. Generally, the occurrence of twinning at low temperatures 
can be understood by the fact that, in BCC metals the Peierls stress for perfect dislocations increases more rapidly with 
decreasing temperature than that for partial dislocations \cite{Hirth}. As a result, the glide of partial dislocations 
(twinning) is easier at low temperatures. However, the twinning to slip transition with decreasing size in ultrathin 
$<$100$>$/\{110\} BCC Fe nanowires is interesting (Figure \ref{Figure4}a) and for the first time such a transition in BCC 
nanowire is being reported. However, a similar twinning to slip transition with decreasing size has been observed in Ti 
microcrystals \cite{HCP-Size}. This size dependence of twinning in Ti has been explained based on the simulated slip model 
\cite{HCP-Size}. It has been suggested that a pole of screw dislocation perpendicular to the slip plane acts as a promoter 
for twin nucleation \cite{HCP-Size}. Below certain critical size, the dislocations are not high enough to activate twinning 
and as a result, the twinning is not observed in sample size lower than 1 $\mu m$ \cite{HCP-Size}. The same model cannot 
be used to explain the observed twinning to slip transition due to the absence of initial dislocations in BCC Fe nanowire. 
However, the size dependence of twinning in ultrathin pristine nanowires may 
arise due to highly coordinated nature of twinning mechanism. It is well known that the twin grows by the systematic glide 
of twinning dislocations having the same Burgers vector on adjacent parallel planes. This coordinated or coherent phenomenon 
may get disturbed in nanowires of less than a certain size, where the number of surface atoms remains higher than the core 
atoms. As a result, the twinning is not observed in ultrathin nanowires. Similarly, increasing the temperature have the 
same effect as that of decreasing the nanowire size, i.e., the high temperature may also disturb the coordinated behaviour 
of twinning phenomenon in ultra-thin nanowires. The size dependence of twinning has also been observed in nanocrystalline 
materials \cite{Twin-grain}. Below certain grain size, it has been reported that the propensity for deformation twinning 
decreases with decreasing grain size \cite{Twin-grain}.

\section{\large Conclusions}

We have performed extensive MD simulations on the tensile deformation of ultrathin $<$100$>$/\{110\} BCC Fe nanowires for 
different sizes and temperatures. BCC Fe nanowires with cross-section width less than 3.23 nm deform by twinning mechanisms 
at low temperatures, while dislocation slip dominates at high temperatures. This indicates that the BCC Fe nanowires undergo 
twinning to slip transition with decreasing size and increasing temperature. Further, the temperature at which the nanowires 
show twinning to slip transition, increases with increase in nanowire size, and above 3.23 nm, deformation twinning dominates 
at all temperatures. The twinning to slip transition has not been observed under the compressive loading of the nanowires.\\

}


\begin{thebibliography}{99}

{\footnotesize

\baselineskip 4pt

\bibitem{Greer-PMS} J.R. Greer, J.Th.M. De Hosson, Prog. Mater. Scie. 56 (2011) 654-724.

\bibitem{Wei-Cai-JMC} C.R. Weinberger, W. Cai, J. Mater. Chem. 22 (2012) 3277-3292.

\bibitem{Liang} H. Liang, M. Upmanyu, H. Huang, Phys. Rev. B 71 (2005) 241403 (R).

\bibitem{Uchic} M.D. Uchic, D.M. Dimiduk, J.N. Florando, W.D. Nix, Science 305 (2004) 986-989.

\bibitem{cross-over-NL} Y. Yue, P. Liu, Q. Deng, E. Ma, Z. Zhang, X. Han, Nano lett. 12 (2012) 4045-4049.

\bibitem{Ag-transition} D. Kong, S. Sun, T. Xin, L. Xiao, X. Sha, Y. Lu, S. Mao, J. Zou, L. Wang, X. Han, J. Alloys Compd. 
676 (2016) 377-382.

\bibitem{Oh} S.H. Oh, M. Legros, D. Kiener, P. Gruber, G. Dehm, Acta Mater. 55 (2007) 5558-5571.

\bibitem{Sedlmayr} A. Sedlmayr, E. Bitzek, D.S. Gianola G. Richter, R. Monig, O. Kraft, Acta Mater. 60 (2012) 3985-3993.

\bibitem{Roos} B. Roos, B. Kapelle, G. Richter, C.A. Volkert, Appl. Phys. Lett. 105 (2014) 201908.

\bibitem{Small} C. Peng, Y. Zhan, J. Lou, Small 8 (2012) 1889-1894.

\bibitem{HCP-Size} Q. Yu, Z.W. Shan, J. Li, X. Huang, L. Xiao, J. Sun, E. Ma, Nature 463 (2010) 335-338.

\bibitem{Mo} P. Wang, W. Chou, A. Nie, Y. Huang, H. Yao, H. Wang, J. Appl. Phys. 110 (2011) 093521.

\bibitem{Healy} C. J. Healy, G. J. Ackland, Acta Mater. 70 (2014) 105-112.


\newpage

\bibitem{Sainath-Orientation} G. Sainath, B. K. Choudhary, Comput. Mater. Sci. 111 (2016) 406-415.

\bibitem{Adutta} A. Dutta, Acta Mater. 125 (2017) 219-230.

\bibitem{Hagen} A. B. Hagen, B. D. Snartland, and C. Thaulow, Acta Mater. 129 (2017) 398-407.


\bibitem{Sainath-100-size} G. Sainath, B. K. Choudhary, T. Jayakumar, Comput. Mater. Sci. 104 (2015) 76-83.

\bibitem{Nature-exp} J. Wang, Z. Zeng, C. R. Weinberger, Z. Zhang, T. Zhu, and S. X. Mao, Nat. Mater. 14 (2015) 594-600. 

\bibitem{LAMMPS} S. J. Plimpton, J. Comput. Phys. 117 (1995) 1-19.

\bibitem{Mendelev} M. I. Mendelev, S. Han, D. J. Srolovitz, G. J. Ackland, D. Y. Sun, M. Asta, Philos. Mag. 83 (2003) 3977-3994.

\bibitem{Virial-Yip} K. S. Cheung and S. Yip, J. Appl. Phys. 70 (1991) 5688-5690.

\bibitem{Virial-Zhou} M. Zhou, Proc. R. Soc. London, Ser. A 459 (2003) 2347-2392.

\bibitem{AtomEye} J. Li, Modell. Simul. Mater. Sci. Eng. 11 (2003) 173-177.

\bibitem{OVITO} A. Stukowski, Modell. Simul. Mater. Sci. Eng. 18 (2010) 015012.

\bibitem{Bulatov} L.A. Zepeda-Ruiz, A. Stukowski, T. Oppelstrup, V.V. Bulatov, Nature 550 (2017) 492-495. 

\bibitem{Sai-PML} G. Sainath, B.K. Choudhary, Phil. Mag. Lett. 96 (2016) 469-76.



\bibitem{Hirth} J.P. Hirth, J. Lothe, Theory of Dislocations, McGraw-Hill, New York, 1968.

\bibitem{Twin-grain} X. L. Wu, Y.T. Zhu, Phys, Rev. Lett. 101 (2008) 025503. 

}

\end{thebibliography}
\end{document}